\documentclass[aps,showpacs,preprint,preprintnumbers,amsmath,amssymb,nofootinbib]{revtex4}
\usepackage{epsfig}
\usepackage{color,graphicx}
\usepackage{dcolumn}
\usepackage{bm}
\usepackage{feynmp}
\usepackage{slashed}

\def\beq{\begin{equation}}
\def\eeq{\end{equation}}

\newcommand{\bea}{\begin{eqnarray}}
\newcommand{\eea}{\end{eqnarray}}

\def\eeqn{\end{equation}}
\newcommand\iden{\leavevmode\hbox{\small1\normalsize\kern-.33em1}}

\let\jnfont=\rm
\def\NPB#1,{{\jnfont Nucl.\ Phys.\ B }{\bf #1},}
\def\PLB#1,{{\jnfont Phys.\ Lett.\ B }{\bf #1},}
\def\EPJC#1,{{\jnfont Eur.\ Phys.\ Jour.\ C }{\bf #1},}
\def\PRD#1,{{\jnfont Phys.\ Rev.\ D }{\bf #1},}
\def\PRL#1,{{\jnfont Phys.\ Rev.\ Lett.\ }{\bf #1},}
\def\MPLA#1,{{\jnfont Mod.\ Phys.\ Lett.\ A }{\bf #1},}
\def\JPG#1,{{\jnfont J.\ Phys.\ G }{\bf #1},}
\def\CTP#1,{{\jnfont Commun.\ Theor.\ Phys.\ }{\bf #1},}
\def\JHEP#1,{{\jnfont JHEP \ }{\bf #1},}
\def\NPPS#1,{{\jnfont Nucl.\ Phys.\ Proc.\ Suppl.\ }{\bf #1},}

\begin{document}

\begin{flushright}
\normalsize{IPMU14-0007}
~\\
\end{flushright}

\title{Electroweak baryogenesis in the MSSM with vector-like superfields}

\author{{Xue Chang}, {Ran Huo}}
\affiliation{
State Key Laboratory of Theoretical Physics, Institute of Theoretical Physics,
Chinese Academy of Sciences, P.O. Box 2735, Beijing 100190, China\\
Kavli IPMU (WPI), The University of Tokyo, 5-1-5 Kashiwanoha, Kashiwa, CHiba 277-8583, Japan}
\email{chxue@itp.ac.cn, ran.huo@ipmu.jp}

\begin{abstract}
Introducing heavy particles with strong couplings to the Higgs field can strengthen
electroweak phase transition, through the entropy release mechanism from both bosons and fermions.
We analyze the possibility of electroweak baryogenesis
in the MSSM with new vector-like superfields. The new vector-like particles belong to the
representation $5+\overline{5}+10+\overline{10}$ of $SU(5)$.
By analyzing in detail the effective potential at finite temperature,
we show that a strongly first order electroweak phase transition in this model is ruled out by a combination of 125 GeV Higgs requirement,
the bound for exotic quarks, the gluon fusion Higgs production rate and the Higgs diphoton decay
rate as well as the electroweak precision measurement.
\end{abstract}


\maketitle

\section{Introduction}

The origin of the matter anti-matter asymmetry of our universe remains unclear.
The three Sakharov conditions  \cite{Shaposhnikov} can be fulfilled in high scale
mechanisms such as leptogenesis  \cite{lep-g1,lep-g2} and Grand Unified Theory (GUT)
baryogenesis  \cite{gut-g1,gut-g2,gut-g3,gut-g4}, but are difficult to test by
electroweak (EW) scale experiments. While electroweak baryogenesis (EWBG)  \cite{ewbg},
relying on weak scale physics, provides an alternative solution which requires a
strongly first-order phase transition (SFOPT) \cite{sfopt}. Unfortunately,
the EW phase transition (EWPT) is too weak in the Standard Model (SM) with large
Higgs mass  \cite{sm-ewpt1,sm-ewpt2} and the CP violation is too small  \cite{sm-cp}.

Extensions of the SM with new EW scale physics can lead to a SFOPT, in all of which
new particles beyond the SM are needed. On the other hand, the ATLAS and CMS collaborations
at the CERN Large Hadron Collider (LHC) reported observation of a SM-like Higgs boson with
mass of $125-126$ GeV \cite{atlas,cms}. If we require the EWBG mechanism to account for the matter anti-matter
asymmetry, the new fields introduced for a SFOPT can induce significant corrections to the SM-like
Higgs mass as well as production and decay rates, which will be strongly constrained.
For example in the Minimal Supersymmetric Standard Model (MSSM), the light stop scenario \cite{lstop-1,lstop-2}
has been severely constrained \cite{lstop-3}.

Based on what physics is responsible for generating the
barrier between the symmetric and broken phases, there are three EWPT model classes
in general \cite{ewpt-class}. In this paper, we focus on the thermally driven case. In addition to the
effect induced by terms cubic in $\phi$ in the bosonic high temperature expansion, the phase transition can be strengthened by introducing heavy
particles with strong couplings to the Higgs fields,
such as the SM extension with TeV Higgsinos, Winos
and Binos \cite{newp,Huo}. 
That is, after the electroweak symmetry breaking (EWSB),
the new particles get Yukawa masses and become heavier,
they approximately decouple from the thermal plasma and transfer their entropy into the thermal
bath. 
In this paper we consider a different model, namely in addition to the MSSM,
adding several vector-like (VL) superfields. This kind of model \cite{vl1,vl2} have been
extensively studied and found interesting, for it can relax the naturalness problem raised by
the Higgs mass, be consistent with gauge coupling
unification and precision EW measurements, and have a rich phenomenology. So it is interesting to
explore its possibility to realize the SFOPT in detail.

The added exotic particles belong to the representation
$5+\overline{5}+10+\overline{10}$ of $SU(5)$,
which consists four new quarks, two new charged leptons, two left handed neutrinos
and the corresponding sparticles with total degree of freedom 120. The model is the MSSM with two new supersymmetric generations,
while VL mass terms are introduced between the two to escape the experimental 4th generation search bound.
In search for a SFOPT we analyze in details the zero temperature potential, the one-loop
zero temperature potential and the finite temperature potential.
To search for viable parameter region we also impose all conventional constraints: the SM like Higgs mass is about $125$ GeV,
no new light quarks of a few hundred GeV exist \cite{PDG}, the gluon fusion Higgs production
rate and the Higgs diphoton decay rate are not significantly changed \cite{atlas,cms}, and the
Peskin-Takeuchi parameters $T$ and $S$ \cite{EWST} are small.

We find generally a SFOPT combining with a 125 GeV Higgs requirement will lead
to a too light exotic fermion/scalar. In order to make them heavy enough to escape the direct
search bound the VL masses should be about $500$ GeV, but the VL Yukawa are also pushed to
large values near the perturbativity bound. We find an almost supersymmetric VL
sector with large $\tan \beta$ and no scalar mixing as our best
solution, which can satisfy the $125$ GeV Higgs requirement
without changing the Higgs gluon fusion rate and the Higgs diphoton decay rate. However, it is
still in tension with the direct light new particle search, and eventually ruled
out by contributing a very large Peskin-Takeuchi $T$ parameter. So in all, the
possibility of EWBG induced by supersymmetric VL generations in our setup is fairly ruled out.

The outline of the rest of the paper is as follows: We will define the model precisely in section II. In section III and IV we investigate
the zero temperature potential (as well as the Higgs mass) and the finite temperature potential separately.
Section V contains our final results and discussions of various constraints. A brief summary is given in
the last section.

\section{The MSSM with Vector-Like Superfields}

As mentioned above, new particles beyond the MSSM are two new generations
$5+\overline{5}+10+\overline{10}$ of $SU(5)$. 
Here we do not take the singlet right hand neutrino into account,
so there will be no Yukawa couplings of the VL neutrinos and the neutrinos do
not contribute to EWSB. Moreover, the model almost preserve
gauge couplings unification \cite{vl-gut}, so it is also UV motivated.

The corresponding quantum numbers of VL superfields under
SU(3)$_c\times$SU(2)$_L\times$U(1)$_Y$ are given as
\begin{equation}
\begin{array}{c}
Q(3, 2,\frac{1}{3}),~~~~~U(3, 1, \frac{4}{3}),~~~~
D(3, 1,-\frac{2}{3}),~~~L(1,2,-1),~~~E(1,1,-2),\\
\overline{Q}(\bar{3}, 2, -\frac{1}{3}),~~~~
\overline{U}(\bar{3}, 1, -\frac{4}{3}),~~~~\overline{D}(\bar{3}, 1, \frac{2}{3}),~~~~
\overline{L}(1, 2,1),~~~~
\overline{E}(1, 1,2).
\end{array}
\end{equation}
And the superpotential is
\begin{equation}
\label{superpotential}
\begin{array}{lll}
W & = & W_{\mathrm{MSSM}}
 + M_Q \overline{Q} Q + M_U \overline{U} U
 + M_D \overline{D} D+ M_L \overline{L} L+ M_E \overline{E} E\\
 &&+ k_{1} H_u Q \overline{U}
 + k_{2} H_u \overline{Q} D
 + k_{3} H_u \overline{L} E
 - k_{1}^{'} H_d \overline{Q} U
 - k_{2}^{'} H_d Q \overline{D}
 - k_{3}^{'} H_d L \overline{E}~.
\end{array}
\end{equation}
Note that in general there are mixing between the new vector-like superfields and the MSSM
superfields. The related Yukawa couplings with the first/second family
MSSM fields are strongly constrained by the EW phenomenology such as flavor
changing neutral current \cite{vl-constrian}, which need to be less than $10^{-3}$. The constraint on the
couplings with the third family MSSM fields is relatively loose, which can be of order
$0.1$. We ignore the effect of these terms just in the EWPT calculation for simplicity.

By assuming universality of the mass-squared terms and the alignment of the B terms,
the soft mass terms and the trilinear soft terms of all the VL scalar partners are given by
\begin{eqnarray}
\label{softbreaking}
-{\mathcal L}_{\mathrm{soft}}&=&m^2_Q|\tilde{Q}|^2+m^2_{\bar{Q}}|\tilde{\bar{Q}}|^2+m^2_U|\tilde{U}|^2
+m^2_{\bar{U}}|\tilde{\bar{U}}|^2+m^2_D|\tilde{D}|^2+m^2_{\bar{D}}|\tilde{\bar{D}}|^2 \nonumber\\
&&+m^2_L|\tilde{L}|^2+m^2_{\bar{L}}|\tilde{\bar{L}}|^2+m^2_E|\tilde{E}|^2+m^2_{\bar{E}}|\tilde{\bar{E}}|^2
+(B_QM_Q \tilde{Q}\tilde{\bar{Q}}+B_UM_U \tilde{U}\tilde{\bar{U}}\nonumber\\
&&+B_DM_D \tilde{D}\tilde{\bar{D}}
+B_LM_L \tilde{L}\tilde{\bar{L}}+B_EM_E \tilde{E}\tilde{\bar{E}}
+A_{k_t}k_1 H_u \tilde{Q} \tilde{\bar{U}}+A_{k_{b}}k_{2} H_u \tilde{\bar{Q}} \tilde{D}\nonumber\\
&&-A_{k_{t}^{'}}k_{1}^{'}H_d \tilde{\bar{Q}} \tilde{U}-A_{k_{b}^{'}}k_{2}^{'}H_d \tilde{Q} \tilde{\bar{D}}
-A_{k_{\tau}^{'}}k_{3}^{'}H_d \tilde{L} \tilde{\bar{E}}
+\mathrm{c.c.}).
\end{eqnarray}

From Eq.~(\ref{superpotential}, \ref{softbreaking}), the new charged fermions field-dependent mass matrices are
\begin{equation}
\label{VLfermionMass}
{\mathcal M}_U(\phi) =
\begin{pmatrix}
M_Q & k_{1} \frac{ \phi_u }{\sqrt{2}} \\
k_{1}^{'} \frac{\phi_d}{\sqrt{2}} & M_U \\
\end{pmatrix}
~,~
{\mathcal M}_D(\phi) =
\begin{pmatrix}
M_Q & k_{2}^{'} \frac{\phi_d}{\sqrt{2}}  \\
k_{2} \frac{ \phi_u }{\sqrt{2}} & M_D \\
\end{pmatrix}
~,~
{\mathcal M}_E(\phi) =
\begin{pmatrix}
M_L & k_{3}^{'} \frac{\phi_d}{\sqrt{2}}  \\
k_{3} \frac{ \phi_u }{\sqrt{2}}  & M_E \\
\end{pmatrix}
.
\end{equation}
We have defined\footnote{In this paper we use
$s_\beta, c_\beta$ for $\sin\beta, \cos\beta$} $\langle\phi_{d}\rangle=v_d=c_\beta v$ and
$\langle\phi_u\rangle=v_u=s_\beta v$ and $v\simeq246$ GeV.
The corresponding field-dependent sfermion squared-mass matrix, for new up type squark for instance, is
\begin{equation}
\label{VLscalarUpMass}
\mathcal{M}^2_{\tilde{U}}=\begin{pmatrix}
m_{\tilde{{t}}_{L^{'}}}^2 & m_{X_{t^{'}}}^2 &B_QM_Q &M^*_Q k_1 \frac{ \phi_u }{\sqrt{2}} +M_Uk'_1\frac{\phi_d}{\sqrt{2}} \\
m_{X_{t^{'}}}^2 & m_{\tilde{{t}}_{R^{'}}}^2 &M_U k_1\frac{ \phi_u }{\sqrt{2}} +M^*_Q k'_1\frac{\phi_d}{\sqrt{2}}& B_UM_U\\
B_QM_Q & M_U k_1\frac{ \phi_u }{\sqrt{2}} +M^*_Q k'_1\frac{\phi_d}{\sqrt{2}} & m_{\tilde{{t}}_{L^{''}}}^2 & m_{X_{t^{''}}}^2\\
M^*_Q k_1 \frac{ \phi_u }{\sqrt{2}} +M_Uk'_1\frac{\phi_d}{\sqrt{2}} & B_UM_U & m_{X_{t^{''}}}^2 & m_{\tilde{{t}}_{R^{''}}}^2
\end{pmatrix},
\end{equation}
in which the basis is $(\bar{Q}^*, U, Q, \bar{U}^*)$, and we have defined
\begin{equation}\label{VLscalarUpMass2}
\begin{array}{lll}
 m_{\tilde{t}_{L'}}^2(\phi)= M_Q^2+m_{\bar{Q}}^2 +\frac{1}{2}k_{1}^{'2} \phi_d^2+D_{\tilde{t}_{L'}}^2(\phi)\\
 m_{\tilde{t}_{R'}}^2(\phi)= M_U^2+m_{U}^2 +\frac{1}{2}k_{1}^{'2} \phi_d^2+D_{\tilde{t}_{R'}}^2(\phi)\\
 m_{\tilde{t}_{L''}}^2(\phi)= M_Q^2+m_{Q}^2 +\frac{1}{2}k_1^2 \phi_u^2+D_{\tilde{t}_{L''}}^2(\phi)\\
 m_{\tilde{t}_{R''}}^2(\phi)= M_U^2+m_{\bar{U}}^2 +\frac{1}{2}k_1^2 \phi_u^2+D_{\tilde{t}_{R''}}^2(\phi)\\
 m_{X_{t'}}^2(\phi)=k_{1}^{'} (A_{k_{t'}}\frac{\phi_d}{\sqrt{2}}-\mu\frac{ \phi_u }{\sqrt{2}})\\
 m_{X_{t''}}^2(\phi)=k_1 (A_{k_{t}}\frac{ \phi_u }{\sqrt{2}}-\mu\frac{\phi_d}{\sqrt{2}})\\
 D_{\tilde{t}_{L'}}^2(\phi)=-D_{\tilde{t}_{L''}}^2(\phi)=-(\frac{g^2}{8}-\frac{g'^2}{12})(\phi_d^2-\phi_u^2),\\
D_{\tilde{t}_{R'}}^2(\phi)=-D_{\tilde{t}_{R''}}^2(\phi)=-\frac{g'^2}{6}(\phi_d^2-\phi_u^2).
\end{array}
\end{equation}
The squared-mass matrices for down type squark and charged slepton are similar.
After diagonalization we get two new Dirac up-type quarks $t'_{1,2}$, two new Dirac down-type quarks $b'_{1,2}$,
two new Dirac charged leptons $\tau'_{1,2}$, and two new
left-handed neutrino $\nu'_{1,2}$ as well as their superpartners
$\tilde{t}'_{1,2,3,4}$, $\tilde{b}'_{1,2,3,4}$, $\tilde{\tau}'_{1,2,3,4}$,\
and $\tilde{\nu}'_{1,2}$\footnote{Strictly speaking (s)neutrinos don't need diagonalization.}.

In the following calculation we neglect all the D-terms and B-terms in the mass matrices
\footnote{At phase transition the D-terms are comparable with top squark thermal mass in Eq.~(\ref{stopMatrix2}) which we are not ignoring, but here we have more important contribution from VL Yukawa couplings in any way.}.
For simplicity we further assume at low scale (namely without renormalization group equation (RGE) running):
\begin{eqnarray}\label{simplify1}
m^2_{Q}=m^2_{\bar{Q}}=m^2_{U}=m^2_{\bar{U}}=m^2_{D}=m^2_{\bar{D}}=m^2_{L}=m^2_{\bar{L}}=m^2_{E}=m^2_{\bar{E}}&=&m^2,\nonumber\\
M_Q=M_{U}=M_{D}=M_{L}=M_{E}&=&M_{V},\nonumber\\
A_{k_{t,b,\tau}}=A_{k_{t',b',\tau'}}&=&A,
\end{eqnarray}
and define the VL scalar squared-mass average and the mass mixing parameter as
\begin{eqnarray}\label{simplify2}
M^2_{S}&=&M^2_{V}+m^2, \nonumber\\
X_{1}&=&A-\mu \mathrm{cot}\beta,\nonumber\\
X_{2}&=&A-\mu \mathrm{tan}\beta.
\end{eqnarray}

We choose $\tan\beta=10$ as our benchmark. Note that the Yukawa $k_{1,2,3}$
are always combined with $\phi_u$ and the Yukawa $k'_{1,2,3}$ with $\phi_d$,
the latter is always suppressed by $\tan\beta$. We actually set $k'_{1,2,3}$ to zero
(see the discussion of the gluon fusion and Higgs diphoton deacy), then $\phi_d$ decouples.
Arising from the first mass matrix in Eq.~(\ref{VLfermionMass}),
the field-dependent squared-mass eigenvalues of $t^{'}_{1,2}$ can be simplified as
\begin{eqnarray}\label{VLfermionMassSolution}
m_{{t^{'}_{1,2}}}^2 (\phi_u,\phi_d)& =&M^2_V+\frac{1}{4}k^2_1 \phi^2_u
 \mp \frac{1}{4}\sqrt{k^4_1 \phi^4_u+8M^2_Vk^2_{1}\phi^2_u }~,
\end{eqnarray}
and the four field-dependent squared-mass eigenvalues, arising from Eq.~(\ref{VLscalarUpMass},\ref{VLscalarUpMass2}), are
\begin{eqnarray}\label{VLscalarMassSolution}
m_{{\tilde{t}^{'}_{1}}}^2 (\phi_u,\phi_d)& =&M^2_S+\frac{1}{4}k^2_1 \phi^2_u-\frac{1}{2\sqrt{2}}k_1 \phi_uX_1\nonumber\\
&&- \frac{1}{2}\sqrt{(\frac{1}{2}k^2_1 \phi^2_u-\frac{1}{\sqrt{2}} k_1 \phi_uX_1)^2+2M^2_Vk^2_{1}\phi^2_u },\\
m_{{\tilde{t}^{'}_{2}}}^2 (\phi_u,\phi_d)& =&M^2_S+\frac{1}{4}k^2_1 \phi^2_u+\frac{1}{2\sqrt{2}}k_1 \phi_uX_1\nonumber\\
&&- \frac{1}{2}\sqrt{(\frac{1}{2}k^2_1 \phi^2_u+ \frac{1}{\sqrt{2}}k_1 \phi_uX_1)^2+2M^2_Vk^2_{1}\phi^2_u },\\
m_{{\tilde{t}^{'}_{3}}}^2(\phi_u,\phi_d)& =&M^2_S+\frac{1}{4}k^2_1 \phi^2_u-\frac{1}{2\sqrt{2}}k_1 \phi_uX_1\nonumber\\
&&+\frac{1}{2}\sqrt{(\frac{1}{2}k^2_1 \phi^2_u- \frac{1}{\sqrt{2}}k_1 \phi_uX_1)^2+2M^2_Vk^2_{1}\phi^2_u },\\
m_{{\tilde{t}^{'}_{4}}}^2(\phi_u,\phi_d)& =&M^2_S+\frac{1}{4}k^2_1 \phi^2_u+\frac{1}{2\sqrt{2}}k_1 \phi_uX_1\nonumber\\
&&+\frac{1}{2}\sqrt{(\frac{1}{2}k^2_1 \phi^2_u+ \frac{1}{\sqrt{2}}k_1 \phi_uX_1)^2+2M^2_Vk^2_{1}\phi^2_u }.
\end{eqnarray}
For field-dependent masses of new down-type quarks, new charged leptons and their superpartners,
one just need to substitute $k_1\rightarrow k_2, k_3$.

At the end of this section, we give the direct search limits on new particles.
As mentioned before, the exotic heavy fermions can decay into SM particles when kinematically
allowed through the mixing Yukawa couplings \cite{vl1,vl2}.
Direct searches set limits to the exotic fermions in such decay modes.
Limits on sparticles depend on the mixing angles of the mass eigenstates
and the mass splittings between them and the lightest neutralino.
The strongest current limits on the extra quarks, leptons and their scalar particles are given as \cite{PDG}
\begin{eqnarray}
&&m_{t'}> 685 \mathrm{GeV} ~,~m_{\tilde{t}'}> 95.7 \mathrm{GeV}, \\
&&m_{b'}> 685 \mathrm{GeV} ~,~m_{\tilde{b}'}> 89 \mathrm{GeV}, \\
&&m_{\tau'} >100.8 \mathrm{GeV} ~,~m_{\tilde{\tau}'}> 81.9 \mathrm{GeV}, \\
&&m_{\nu'} >39.5 \mathrm{GeV} ~,~m_{\tilde{\nu}'}> 94 \mathrm{GeV}.
\end{eqnarray}
However when considering various combinations of decay modes of new fermions
and not being limited to a special mass constrain for scalars, the above bounds are relaxed.
We will see later that the mass of charged exotic fermions is important to
an acceptable SFOPT, so here in our work we consider some optimistic mass limits
for new charged fermions. Namely we consider $m_{t'}> 415 \mathrm{GeV}$ for $t'$ \cite{T}, which is achieved
by scanning the exotic decay branching ratio triangle,
and $m_{b'}> 360 \mathrm{GeV}$ \cite{vl1} for $b'$ and
$m_{\tau'}> 63.5 \mathrm{GeV}$ \cite{tau} for $\tau'$.
The mass limits for other new particles still take the values shown above.

\section{Zero Temperature Potential and Higgs Mass}

In this model, the zero temperature effective potential at one-loop level are given by
\begin{eqnarray}
V(\phi_u,\phi_d, T=0)&=&V_0(\phi_u,\phi_d)+V_1(\phi_u,\phi_d)
\end{eqnarray}
in which $V_0$ is the tree-level potential,
$V_1$ is the zero-temperature renormalized one-loop potential.

\subsection{Tree Level Potential}
The zero temperature tree-level potential here in our model is the same as in the MSSM, which
is given as
\begin{eqnarray}
V_{\mathrm{MSSM}}&=&\frac{1}{2}m^2_{11}\phi_d^2+\frac{1}{2}m^2_{22}\phi_u^2-m^2_{12}\phi_d\phi_u
+\frac{1}{4}\lambda_1\phi_d^4+\frac{1}{4}\lambda_2\phi_u^4
+\frac{1}{2}\lambda_3\phi_d^2 \phi_u^2,
\end{eqnarray}
in which
\begin{eqnarray}
m^2_{11}&=&m^2_{H_d}+\mu^2,\\
m^2_{22}&=&m^2_{H_u}+\mu^2,\\
m^2_{12}&=&b\mu,\\
\lambda_1&=&\lambda_2=-\lambda_3=\frac{1}{8}(g^2+g'^{2}).
\end{eqnarray}

\subsection{The Renormalization
Group Improved Higgs Potential and the SM-Like Higgs Mass}

The third generation MSSM particles and the new VL particles will induce significant
corrections to the Higgs potential. Here we are interested in the complete one loop improved
Higgs potential, because it determines the SM like Higgs mass. We follow  \cite{Haber} to write it as
\begin{eqnarray}\label{V0inProgram}
V_{\mathrm{MSSM}}&=&\frac{1}{2}(m^2_{11}+\Delta m^2_{11})\phi_d^2+\frac{1}{2}(m^2_{22}+\Delta m^2_{22})\phi_u^2-(m^2_{12}+\Delta m^2_{12})\phi_d\phi_u\nonumber\\
&&+\frac{1}{4}(\lambda_1+\Delta\lambda_1)\phi_d^4+\frac{1}{4}(\lambda_2+\Delta\lambda_2)\phi_u^4
+\frac{1}{2}(\lambda_3+\Delta\lambda_3)\phi_d^2\phi_u^2\nonumber\\
&&+\Delta\lambda_6\phi_d^3\phi_u+\Delta\lambda_7\phi_d\phi_u^3,
\end{eqnarray}
where $\Delta\lambda_6  \phi_d^3 \phi_u $ and $\Delta\lambda_7 \phi_d \phi_u^3 $ are the one-loop potential induced
terms which don't exist in the tree-level potential.
The expressions for the corrections are listed in Appendix A.

With the renormalization
group (RG) improved Higgs potential, the SM-like Higgs mass can be written as
\begin{eqnarray}
m^2_{h_0}&=&m^2_Z \cos^2 2\beta+2\Delta\lambda_1 v^2\sin^4 \beta+2\Delta\lambda_2 v^2\cos^4 \beta
+4\Delta\lambda_3 v^2\sin^2 \beta \cos^2 \beta \nonumber\\
&&+8\Delta\lambda_6 v^2\sin \beta \cos^3 \beta+8\Delta\lambda_7 v^2\sin^3 \beta \cos \beta.
\end{eqnarray}
In order to get a simple analytical expression, we set the parameters as mentioned before
and further set
\begin{eqnarray}\label{simplify3}
k_1=k_2=k_3=k~,~
\end{eqnarray}
then 
the SM-like Higgs mass can be simplified as
\begin{eqnarray}
m^2_{h_0}&=&m^2_Z \cos^2 2\beta+\frac{3v^2}{4\pi^2}y_t^4
[\mathrm{ln}\Big(\frac{\tilde{m}_t}{m_t}\Big)+\frac{X^2_t}{2\tilde{m}^2_t}\Big(1-\frac{X^2_t}{12\tilde{m}^2_t}\Big)]\\
&&+\frac{7v^2}{8\pi^2} k^4s^4_\beta\Bigl [\mathrm{ln}\frac{M^2_S}{M^2_V}-\frac{1}{6}\Big(5-\frac{M^2_V}{M^2_S}\Big)\Big(1-\frac{M^2_V}{M^2_S}\Big)
+\hat{X}^2_{1}\Big(1-\frac{M^2_V}{3M^2_S}\Big)-\frac{\hat{X}^4_{1}}{12}\Bigr ].
\nonumber
\end{eqnarray}
We can see that new heavy particles give extra contributions and permitting relatively lighter stop mass, which can loose the tension of the naturalness problem.

\subsection{Zero Temperature One-loop Level Potential}

In the above analysis we actually run the RGE top down from the supersymmetry breaking scale, in order to fix the low energy Higgs mass to be the observed value. However, as we go to higher scales where the EW phase transition takes place, the RGE running is backwards from the low energy potential Eq.~(\ref{V0inProgram}). We describe this process in the way of (zero temperature) one loop potential, which is equivalent to RGE\footnote{We choose to present the one-loop issue in this awkward way because this is the way we do the numerical work: the Coleman-Weinberg form one loop potential are always implemented by a build-in function in the code CosmoTransition \cite{CosmoTransition}, so the low scale parameters consistent with Higgs mass and VEV need to be run down from the supersymmetry breaking scale at first.}. The zero-temperature one-loop potential are given by
\begin{eqnarray}
V_1(\phi_u,\phi_d)=\frac{1}{64\pi^2}\sum_{i}n_im^4_i(\phi_u,\phi_d)\left[~\mathrm{log}\frac{m^2_i(\phi_u,\phi_d)}{Q^2}-c_i~\right]
\end{eqnarray}
where $m_i(\phi_u,\phi_d)$ are the field-dependent masses and $Q$ is the
renormalization scale\footnote{A variation of $Q$ induces variation of $\phi^2$ and $\phi^4$ terms,
in Eq.~(\ref{125and246}-\ref{WithCounterTerm}) we see that the combination of them together with counterterms are determined by the renormalization condition, so the value of $Q$ is immaterial.}.
$i$ stands for the particles which can contribute to the
effective potential, $n_i$ is the particle degree of freedom, $c_i$'s are constants which are 5/6 for gauge bosons
and 3/2 for fermions and scalars.
In our work we include the large one-loop corrections
induced by top, stop and all the vector-like
particles as well as the EW gauge bosons, the corresponding degree of freedoms are:
$n_t=n_{t'_{1,2}}=n_{b'_{1,2}}=3 n_{\tau'_{1,2}}=-12$,
$n_{\tilde{t}_{1,2}}=n_{\tilde{t}'_{1,2,3,4}}=n_{\tilde{b}'_{1,2,3,4}}=3n_{\tilde{\tau}'_{1,2,3,4}}=6$,
$n_{W_L}=2$, $n_{W_T}=4$, $n_{Z_L}=1$, $n_{Z_T}=2$, where subscripts L and T means longitudinal and transverse modes respectively.

As stressed above, the one-loop potential should be renormalized in a way which preserves the low energy Higgs VEV and the Higgs mass.
In the one loop potential language it is easy to implement, namely by requiring
\begin{eqnarray}\label{125and246}
\left(\frac{\partial}{\partial \phi_u},\frac{\partial}{\partial \phi_d},\frac{\partial^2}{\partial \phi_u^2},
\frac{\partial^2}{\partial \phi_d^2},\frac{\partial^2}{\partial \phi_u \partial\phi_d}\right)
\big(V_1(\phi_u,\phi_d)+V_1^{\mathrm{c.t.}}(\phi_u,\phi_d)\big)\bigg|_{\phi_d = v_d,\phi_u = v_u}=0.
\end{eqnarray}
Here we introduce the finite ``counterterms'' $V_1^{\mathrm{c.t.}}$ to protect the one-loop potential from shifting the Higgs VEV and CP even Higgs mass matrix. We have five equations so that we can determine up to five coefficients of the counterterm polynomial, here we choose them to be
\begin{eqnarray}
V_1^{\mathrm{c.t.}}=\frac{1}{2}\delta m^2_{11}\phi_d^2+\frac{1}{2}\delta m^2_{22}\phi_u^2
+\frac{1}{4}\delta\lambda_1\phi_d^4+\frac{1}{4}\delta\lambda_2\phi_u^4
+\frac{1}{2}\delta\lambda_3\phi_d^2 \phi_u^2.
\end{eqnarray}
And the corresponding total zero temperature one-loop potential is
\begin{eqnarray}\label{WithCounterTerm}
&&V_1^{\mathrm{re}}(\phi_u,\phi_d)_i=V_1(\phi_u,\phi_d)_i+V_1^{\mathrm{c.t.}}(\phi_u,\phi_d)_i\nonumber\\
&=&\frac{ n_i}{64 \pi^2}\Bigl[m^4_i(\phi_u,\phi_d)\log \frac{m^2_i(\phi_u,\phi_d)}{Q^2}
+\alpha^u_{i} \phi^2_u+\alpha^d_{i}\phi^2_d
+\beta^u_{i} \phi^4_u+\beta^d_{i}\phi^4_d+2\beta^{ud}_{i} \phi^2_u \phi^2_d\Bigr]
\end{eqnarray}

The solution of Eq.~(\ref{125and246}) is unique, namely
\begin{eqnarray}
\alpha^u_{i}&=&(-\frac{3}{2}\frac{\omega_i \omega^{u'}_i}{v_u}+\frac{1}{2}\omega_i^{u'2}+\frac{1}{2}\omega_i \omega^{u''}_i)\log \frac{\omega_i}{Q^2}
-\frac{3}{4}\frac{\omega_i \omega^{u'}_i}{v_u}+\frac{3}{2}\omega_i^{u'2}+\frac{1}{2}\omega_i \omega^{u''}_i-\beta^{ud}_{i}v_d^2\\
\alpha^d_{i}&=&(-\frac{3}{2}\frac{\omega_i \omega^{d'}_i}{v_d}+\frac{1}{2}\omega_i^{d'2}+\frac{1}{2}\omega_i \omega^{d''}_i)\log \frac{\omega_i}{Q^2}
-\frac{3}{4}\frac{\omega_i \omega^{d'}_i}{v_d}+\frac{3}{2}\omega_i^{d'2}+\frac{1}{2}\omega_i \omega^{d''}_i-\beta^{ud}_{i}v_u^2\\
\beta^u_{i}&=&\frac{1}{v^2_u}\bigl[(\frac{1}{4}\frac{\omega_i \omega^{u'}_i}{v_u}-\frac{1}{4}\omega_i^{u'2}-\frac{1}{4}\omega_i \omega^{u''}_i)\log \frac{\omega_i}{Q^2}
+\frac{1}{8}\frac{\omega_i \omega^{u'}_i}{v_u}-\frac{3}{8}\omega_i^{u'2}-\frac{1}{8}\omega_i \omega^{u''}_i\bigr]\\
\beta^d_{i}&=&\frac{1}{v^2_d}\bigl[(\frac{1}{4}\frac{\omega_i \omega^{d'}_i}{v_d}-\frac{1}{4}\omega_i^{d'2}-\frac{1}{4}\omega_i \omega^{d''}_i)\log \frac{\omega_i}{Q^2}
+\frac{1}{8}\frac{\omega_i \omega^{d'}_i}{v_d}-\frac{3}{8}\omega_i^{d'2}-\frac{1}{8}\omega_i \omega^{d''}_i\bigr]\\
\beta^{ud}_{i}&=&-(\frac{\omega^{u'}_i \omega^{d'}_i+\omega_i\omega^{ud''}_i}{4v_u v_d} )\log \frac{\omega_i}{Q^2}
+\frac{3\omega^{u'}_i \omega^{d'}_i+\omega_i\omega^{ud''}_i}{2v_u v_d}
\end{eqnarray}
where we define $\omega_i=m^2_i(v_u,v_d)$,
$\left.\omega^{u(d)'}_i=\frac{\partial m^2_i(\phi_u,\phi_d)}{\partial \phi_{u(d)}}\right|_{ (v_u, v_d)}$,
$\left.\omega^{u(d)''}_i=\frac{\partial^2 m^2_i(\phi_u,\phi_d)}{\partial^2 \phi_{u(d)}}\right|_{ (v_u, v_d)}$
and
$\left.\omega^{ud''}_i=\frac{\partial^2m^2_i(\phi_u,\phi_d)}{\partial \phi_{u}\partial \phi_{d}}\right|_{ (v_u, v_d)}$.
These are the generalization of expressions in \cite{newp} to the two-Higgs doublet model.

\section{Finite Temperature Potential}

The temperature dependent potential at one-loop level are given by
\begin{eqnarray}
\Delta V(\phi_u,\phi_d, T)=\Delta V_1(\phi_u,\phi_d, T)+\Delta V_{\mathrm{daisy}}(\phi_u,\phi_d, T)
\end{eqnarray}
where $\Delta V_1$ is the finite temperature one-loop potential \cite{sm-ewpt2}, and $\Delta V_{\mathrm{daisy}}$
is the finite-temperature effect coming from the resummation of the leading infrared-dominated higher-loop
contributions \cite{sm-ewpt1}. The specific formulas are
\begin{eqnarray}
\Delta V_1(\phi_u,\phi_d,T)&=&\frac{T^4}{2\pi^2}\Bigl \lbrace\sum_{i=\mathrm{bosons}}n_i J_B\Bigl [ \frac{\bar{m}_i(\phi_u,\phi_d)}{T} \Bigr ]
 + \sum_{i=\mathrm{fermions}}n_iJ_F\Bigl [ \frac{m_i(\phi_u,\phi_d)}{T}\Bigr ] \Bigr\rbrace, \\
\Delta V_{\mathrm{daisy}}(\phi_u,\phi_d,T)&=&-\frac{T}{12}\sum_{i=\mathrm{bosons}}
n_i\Bigl [\bar{m}^3_i(\phi_u,\phi_d,T)-m^3_i(\phi_u,\phi_d)
\Bigr ],
\end{eqnarray}
with definitions and high temperature expansions
\begin{eqnarray}\label{JBJF}
J_B\Bigl[ \frac{m}{T} \Bigr]&=&\int^\infty_0 dx~ x^2 \mathrm{log }\Bigl[1-e^{-\sqrt{x^2+\frac{m^2}{T^2}}}\Bigr] = -\frac{\pi^4}{45}+\frac{\pi^2}{12}\frac{m^2}{T^2}-\frac{\pi}{6}\Big(\frac{m^2}{T^2}\Big)^{\frac{3}{2}}+\mathcal{O}\Big(\frac{m^4}{T^4}\Big), \\
J_F\Bigl[ \frac{m}{T} \Bigr]&=&\int^\infty_0 dx~ x^2 \mathrm{log }\Bigl[1+e^{-\sqrt{x^2+\frac{m^2}{T^2}}}\Bigr] =
\frac{7\pi^4}{360}-\frac{\pi^2}{24}\frac{m^2}{T^2}+\mathcal{O}\Big(\frac{m^4}{T^4}\Big),
\end{eqnarray}
in which the thermal mass $\bar{m}^2_i(\phi_u,\phi_d,T)=m^2_i(\phi_u,\phi_d)+\Pi_i(T) $ and $\Pi_i(T)$ is the leading $T$ dependent
self-energy. To leading order, only bosons receive thermal mass corrections. Only the longitudinal components of $W$ and $Z$ receive the daisy corrections.

The thermal masses of the MSSM particles are well known. For the EW gauge bosons the field and temperature dependent masses are
\begin{eqnarray}
m_W^2 (\phi_u,\phi_d,T)&=& \frac{1}{2}g^2(\phi^2_d+\phi^2_u) + \Pi_{W^{\pm}}~,\\
\mathcal{M}^2_{Z\gamma}(\phi_u,\phi_d,T) &=& \left(
\begin{array}{cc}
\frac{1}{2}g^2(\phi^2_d+\phi^2_u) + \Pi_{W^3}&-\frac{1}{2}gg'(\phi^2_d+\phi^2_u)\\
-\frac{1}{2}gg'(\phi^2_d+\phi^2_u)&\frac{1}{2}{g'}^2(\phi^2_d+\phi^2_u) + \Pi_{B}
\end{array}
\right),
\end{eqnarray}
in which the thermal masses $\Pi_{W^{\pm}} = \Pi_{W^3} = \frac{9}{2}g^2T^2$ and $\Pi_{B}=\frac{9}{2}g'^2T^2$
for the longitudinal modes, and for the transverse modes all the thermal masses are zeros.
The field and temperature dependent mass of the MSSM 3rd generation stops are given by
\begin{eqnarray}\label{stopMatrix}
\mathcal{M}_{\tilde{t}}^2 (\phi_u,\phi_d,T)&=&
\left(
\begin{array}{cc}
M_t^2 + \frac{1}{2}y_t^2\phi^2_u
+ \Pi_{\tilde{t}_L} &\frac{1}{\sqrt{2}}y_t\phi_u X_1\\
\frac{1}{\sqrt{2}}y_t\phi_u X_1
&M_t^2 + \frac{1}{2}y_t^2\phi^2_u
+ \Pi_{\tilde{t}_R}
\end{array}
\right),
\end{eqnarray}
where
\begin{eqnarray}\label{stopMatrix2}
\Pi_{\tilde{t}_L} &=& \frac{2}{3}g_s^2T^2 + \frac{1}{72}g^2T^2
+ \frac{3}{8}g'^2T^2 + \frac{1}{4}y_t^2T^2,\\
\Pi_{\tilde{t}_R} &=& \frac{2}{3}g_s^2T^2 + \frac{2}{9}g'^2T^2 + \frac{1}{2}y_t^2T^2.
\end{eqnarray}
All the thermal mass are derive from  Ref. \cite{thermal mass}.
On the other hand, all the new VL particles' thermal masses are neglected
in our work, for both simplicity and nonexistence in literature.
If included, naively it will further rise an order of $g_s^2T^2$ or $k^2T^2$ contribution to the $M_S^2$ terms,
which is probably large and makes a SFOPT even more difficult according to the following discussion.

We calculate the thermal functions $J_{B/F}$ in Eq.~(\ref{JBJF}) numerically instead
of using a high temperature expansion, which is crucial for our purpose. The change in $J_{B/F}$ include the information
of continuous variation of entropy density induced by the new VL particles,
see Fig.~2 in the next section.

\section{Results and Discussion}

We use the public code CosmoTransition  \cite{CosmoTransition} for a numerical evaluation of the phase
transition and perform several scans of the parameter space.
As mentioned before we choose $\tan\beta=10$, we also choose CP odd
Higgs mass $m_A=2000$ GeV for a typical decoupling Higgs sector. For the MSSM
top/stop sector we want a small contribution to the SM like Higgs mass (so that
large contribution from the VL sector and large VL Yukawa coupling are possible),
so we choose $M_{t_L}=700$ GeV, $M_{t_R}=500$ GeV and $A_t=500$ GeV for the soft
breaking parameters and $\mu=500$ GeV. However our results are not sensitive to
the values of the MSSM parameters, because with $X=0$ we can (as we actually do)
choose arbitrarily degenerated fermions and sfermions, $\frac{M_S^2}{M_V^2}\rightarrow1$,
 to reduce their contribution to Higgs mass through the factor
\begin{equation}
\label{factor}
\log\left(\frac{M_S^2}{M_V^2}\right)-\frac{1}{6}\left(5-\frac{M_V^2}{M_S^2}\right)\left(1-\frac{M_V^2}{M_S^2}\right)
\rightarrow
\frac{1}{3}\left(\frac{M_S^2}{M_V^2}-1\right)
\end{equation}

As for the VL parameters, for simplicity in all our scans we set the parameters
as mentioned in Eq.~(\ref{simplify1},\ref{simplify2},\ref{simplify3}), and all the
new down-type Yukawa couplings $k'$ and the down-type mass mixing parameter $X_2$ are taken to be zero.
In scan we have checked that the up-type mass mixing parameter $X_1$ prefers zero in order
to have larger phase transition strength, so we also fix $X_1=0$, which also reduce other contributions
to add to the factor in Eq.~(\ref{factor}) to enable a large Yukawa.

\subsection{SFOPT}

In Fig.~1 we show two scans of phase transition strength with the Yukawa coupling $k$ and the VL mass $M_V$. We also show the constraint of Higgs mass $m_{h^0}\sim 124-127$ GeV and the
lightest new fermion mass contours. On the left panel we fix
$M_S/M_V=1.5$. We can see for such a range of Higgs mass, the SFOPT can
only be achieved for $k\simeq 1.6$ and VL mass $M_V \lesssim 100$ GeV.
On the right panel we fix $M_S/M_V=1.1$, the combination of SFOPT with about 125 GeV Higgs
can only be generated for $k\simeq 2.6$ and VL mass $M_V \lesssim 230$ GeV.
\vspace{2pt}
\begin{figure}[htbp]
\centering
\includegraphics[width=3in]{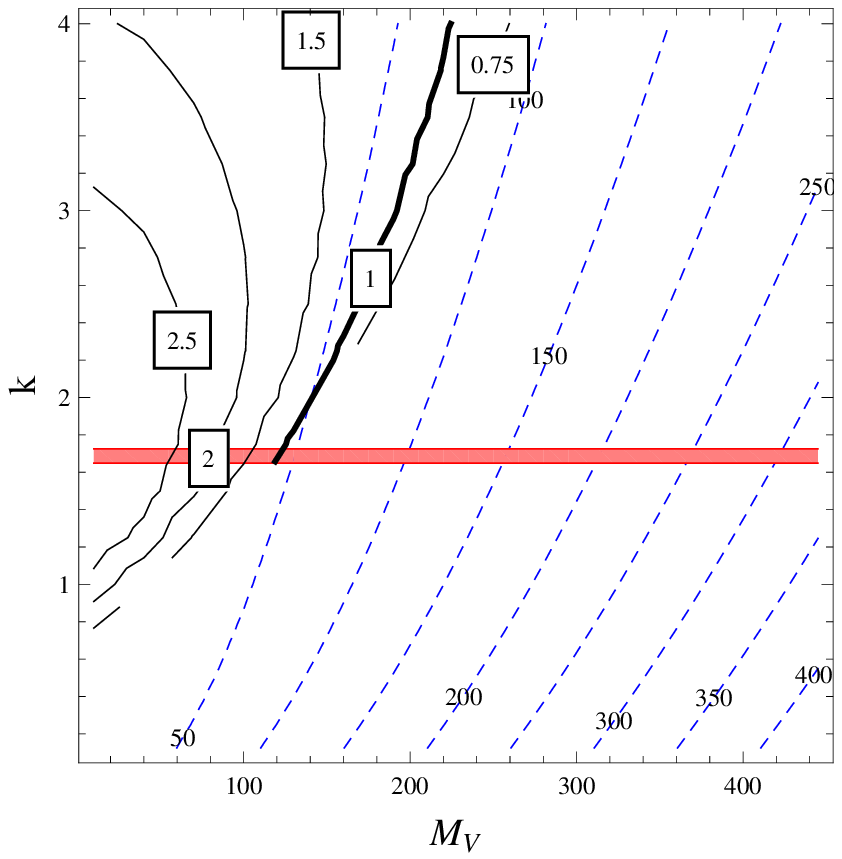}~~~~
\includegraphics[width=3in]{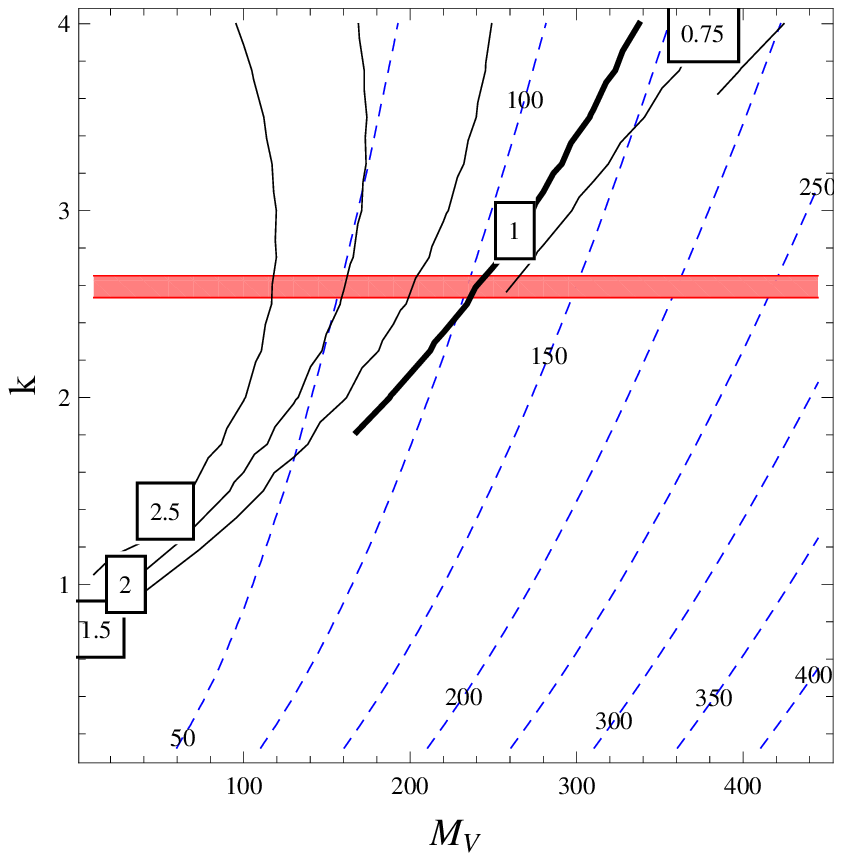}~~~~
\caption{EWBG, Higgs mass and the lightest new fermion mass contours in the MSSM extension with vector-like superfields.
Black curves are the EWPT strength $\phi_c/ T_c$, and blue dashed curves are the lightest new fermion mass. Pink
band is the SM-like Higgs mass region $124-127$ GeV. In the left panel
we fix $M_S/M_V=1.5$, in the right panel we fix $M_S/M_V=1.1$.}
\end{figure}

First we can see, as far as the SFOPT is concerned,
the larger the VL mass $M_V$ is taken, the larger the
Yukawa coupling $k$ needs to be. Because Boltzmann suppression effect
of a few hundred GeV $M_V$ may decouple the new particle in the symmetric
phase, significant entropy release effects for a SFOPT can only be guaranteed by
a large Yuwaka mass and a large $m(\phi)/T$ shift.

Comparing to the entropy release effect in \cite{newp}, we can see that for a SFOPT our required degree of
freedom is much larger\footnote{We note a convention difference and our $k=4$ corresponds to $h=2$ in \cite{newp}.}. This is quantitatively the most
significant point of our analysis. To see clearly, with Eq.~(16-20)
we can write the new fermion mass squares
as $M_{f_{1,2}}^2=M_V^2+\frac{1}{4}k^2 \phi_u^2 \mp \frac{1}{4}\sqrt{k^4 \phi^4_u+8M^2_Vk^2\phi^2_u}$, or equivalently
\begin{equation}
M_{f_{1,2}} = \sqrt{M_V^2+\frac{1}{8}k^2\phi_u^2}\mp\frac{1}{2\sqrt{2}}k\phi_u.
\end{equation}
The new sfermion mass have a similar behavior.
We can understand in the following interesting picture. After EWSB the fermion masses
jump from $M_V$ to $\sqrt{M_V^2+\frac{1}{4}k^2 \phi_u^2}$, and on this basis become split.
The mass splitting terms $\frac{1}{2\sqrt{2}}k\phi_u$
make half of the VL fermions lighter than those in the symmetric phase, overcoming the common shift
$M_V \to \sqrt{M_V^2+\frac{1}{4}k^2 \phi_u^2}$, while the other half heavier.
In Fig.~2 we show the fully calculated finite temperature potential contribution $J_{B/F}$ instead of
only the hight temperature expansions. We can refer to the $J_B, J_F$ curves to see the potential change.
\vspace{2pt}
\begin{figure}[htbp]
\centering
\includegraphics[width=4in]{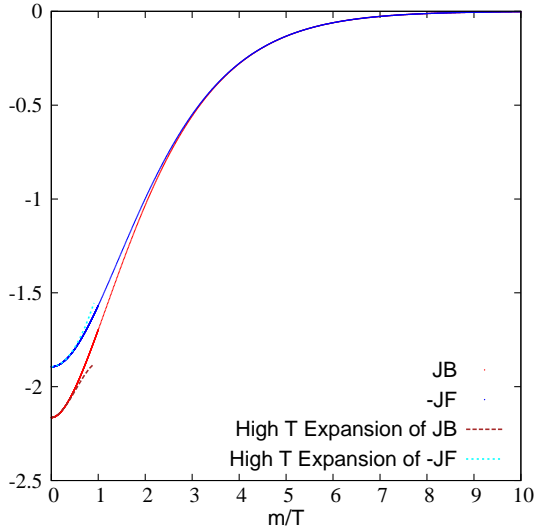}~~~~
\caption{The complete thermal one-loop potential contribution JB (the full lower red curve) and JF (the full upper blue curve) as defined in Eq. (40) and the comparison with their high-temperature expansions (the brown dashed and cyan dotted curves, respectively).
}
\end{figure}

A shift of $M_V/T \to \sqrt{M_V^2+\frac{1}{4}k^2 \phi_u^2}/T$ is exactly the entropy release
effect in \cite{newp}, with an effect of the representative point rise on the $J_F$ curve, or
the thermal potential rise. Here the further new splitting of $\frac{1}{2\sqrt{2}}k\phi_u$ for the heavy
particle will raise more the $m(\phi,T)/T$ and release more entropy, while unfortunately, the
$-\frac{1}{2\sqrt{2}}k\phi_u$ for the light particle will have an opposite effect. A little bit more
quantitative analysis indicates, because the slope of the $J_B/J_F$ curve is less at higher $m/T$
(for example, 4) than at lower $m/T$ (say, 1), the backward splitting $-\frac{1}{2\sqrt{2}}k\phi_u$ to
lower masses always induce a larger thermal potential drop $\Delta J_B/\Delta J_F$ than the forward
splitting $+\frac{1}{2\sqrt{2}}k\phi_u$, and the net effect is a drop, unable to trigger the
SFOPT  \cite{Huo}. This opposite effect will significantly compensate the
$M_V/T \to \sqrt{M_V^2+\frac{1}{4}k^2 \phi_u^2}/T$ effect, making the contribution to phase transition
strength in our scenario much smaller than that with merely the same degree of freedom, the same soft mass
and the same Yukawa, but without splitting.
We will give a more general analysis in our next paper.

\subsection{Higgs Mass and Light Exotic Particle Constraints}

Apparently with SFOPT requirement the first two scans always give too light a new fermion,
so they are ruled out. As we have already discussed, the direction we can go is to increase
$M_V$ and $k$. In Ref. \cite{vl1} an infrared quasi fix point is pointed out, as $k\simeq1.0$
and $h\simeq1.2$. Here we ignore this bound, but the bottom line is the perturbativity bound
 $k\lesssim4$. We choose to saturate the bound, then we get an almost unbroken supersymmetric
 VL sector\footnote{With our MSSM parameter choices we get $\frac{M_S}{M_V}=1.019$.}, see Fig. 3

\vspace{2pt}
\begin{figure}[htbp]
\centering
\includegraphics[width=3in]{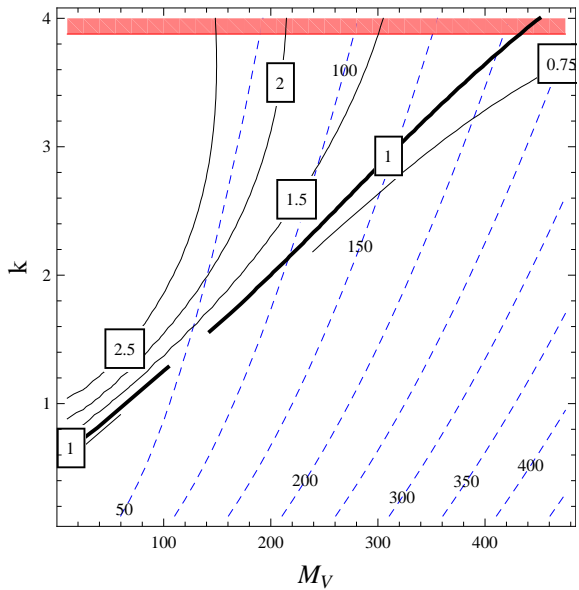}~~~~
\caption{Same as in Fig.~(1), but for $M_S/M_V=1.019$.}
\end{figure}

\begin{table}
\begin{tabular}[c]{|c|c|c|c|c|c|c|c|c|c|c|c|}
\hline
\phantom{i}$M_V$ \phantom{i}&
\phantom{i}$M_S$ \phantom{i}&
\phantom{i}$k$ \phantom{i}&
\phantom{i}$m_{f'_1} $  \phantom{i}&
\phantom{i}$m_{f'_2}$ \phantom{i} &
\phantom{i}$m_{\tilde{f}'_1}$ \phantom{i} &
\phantom{i}$m_{\tilde{f}'_2}$ \phantom{i} &
\phantom{i}$m_{\tilde{f}'_3}$ \phantom{i} &
\phantom{i}$m_{\tilde{f}'_4}$ \phantom{i} &
\phantom{i}$\phi_c/ T_c$ \phantom{i} &
\phantom{i}$m_{h_0}$ \phantom{i}\\ \hline
70 &  105 & 1.6 &17&293 &80 & 80 &304 &304 &2.1 &126.5\phantom{i} \\ 
100 &  150 & 1.6 &32&309 &116 & 116 &329 &329 &1.46  &126.9\\ 
230 &  253 & 2.4 &102&517 &116 & 116 &528 &528 &0.93  &125.7\\ 
475 &  484 & 4.0 &241&934 &259 &259 &939 &939 &0.94 &126.0\\ \hline
\end{tabular}
\caption{Input parameters which can realize both the SFOPT and 124-127 GeV Higgs,
and the corresponding new particle masses.}
\end{table}

The best lightest fermion mass we get is about $241$ GeV, which is still generally
ruled out by heavy $t'$ and $b'$ quark searches, even by optimistic bounds, as mentioned
in Sec. II. We will not discuss the possibility of aligned Yukawa matrix in generation basis,
which make the decay mode nonstandard. On the other hand, the possibility is to relax
the degeneracy between the quarks and the leptons, to make the quark sector $M_V$ and
$M_S$ larger to accommodate heavier new quarks. However at first it is naively against our model
assumption of $5+\overline{5}+10+\overline{10}$ of $SU(5)$ GUT, which predicts $M_Q=M_U=M_E$
and $M_D=M_L$ at the GUT scale. Further we numerically find that due to large zero temperature
corrections
, for separate quark and lepton (or generally two sets)
corrections the potential usually don't even run away from symmetric phase even at zero temperature.
So we will not go into detail of that possibility.

\subsection{Gluon Fusion and Higgs Diphoton Decay Constraints}

We use the low energy theorem \cite{Low} for an estimation. The contributions to the loop amplitude are
all proportional to $\frac{\partial}{\partial\ln v}\det\mathcal{M}$ where $\mathcal{M}$
is any of the mass matrix in Eq.~(\ref{VLfermionMass},\ref{VLscalarUpMass}). As can be see clearly in the
fermion mass matrix, setting all $k'$s to be zeros eventually makes all the
determinants independent of the Higgs VEV. The two masses of Eq.~(\ref{VLfermionMassSolution})
are actually from the matrix $\mathcal{M}\mathcal{M}^\dag$ or $\mathcal{M}^\dag\mathcal{M}$, and $\det(\mathcal{M}\mathcal{M}^\dag)=m_{{t^{'}_1}}^2m_{{t^{'}_2}}^2=M_V^4$ is independent of
$\phi_u,\phi_d$. With $X_1=X_2=0$ the scalar sector has a similar behavior, but there is a
residual contribution proportional to supersymmetry breaking soft parameter $m^2=M_S^2-M_V^2$, namely $\det\mathcal{M}^2_{\tilde{U},\tilde{D},\tilde{E}}=(M_S^4+\frac{1}{2}m^2k^2\phi_u^2)^2$.
Since we are interested in an almost supersymmetric VL sector and $m^2\rightarrow0$,
the corrections to gluon fusion and Higgs diphoton decay amplitudes also vanish in this
limit. So the gluon fusion Higgs production rate and Higgs diphoton decay rate are not
affected. This discussion also justifies our parameter choices $k'=0$, $X_1=0$ and $X_2=0$.

\subsection{Peskin-Takeuchi $T$ and $S$ parameters}

We perform a numerical calculation. The fermionic contribution agrees with the formulas in  \cite{vl1}
\begin{eqnarray}
\Delta T&=&\frac{N_c}{480\pi s_W^2M_W^2M_V^2}\left(\frac{13}{4}(k^4v_u^4+k'^4v_d^4)+\frac{1}{2}(k^3k'v_u^3v_d+k'^3kv_d^3v_u)+\frac{9}{2}k^2k'^2v_u^2v_d^2\right),\\
\Delta S&=&\frac{N_c}{30\pi M_V^2}\left(2(k^2v_u^2+k'^2v_d^2)+kk'v_uv_d\Big(\frac{3}{2}+10Y_\Phi\Big)\right),
\end{eqnarray}
with $Y_\Phi=-\frac{1}{3}$ for our model, while the scalar part nearly gives the same contribution for nearly unbroken supersymmetry. In particular for the last point in Tab.~(1) we get $T\simeq32.5$ and $S\simeq0.2$, which apparently makes it excluded. Such a large $T$ parameter contribution is because it is proportional to $k^4$, and only suppressed by $M_V^2$. On the other hand, the form of the superpotential Eq.~(\ref{superpotential}) determines the custodial symmetry is violated in the maximal way, namely a light left hand up quark component always find a heavy left hand down quark component.

\section{Summary}

We have discussed EWBG in the MSSM extension with vector-like superfields
belonging to the representation $5+\overline{5}+10+\overline{10}$ of $SU(5)$ in detail.
We find the SFOPT has been ruled out by a combination of 125 GeV Higgs requirement,
the direct search for the exotic fermions, the gluon fusion rate and the Higgs diphoton decay rate as well as the EW precision measurement.
However, the general contribution from a (nearly) supersymmetric sector to SFOPT with minimal effect to Higgs phenomenology is still interesting.

\begin{acknowledgments}
We would like to thank Chun Liu and Yu-Feng Zhou for a very helpful discussion.
The work of X.C.~was supported in part by the National Natural Science
Foundation of China under nos.11375248 and 10821504, and by the National
Basic Research Program of China under Grant No. 2010CB833000. The work of
R.H.~was supported by World Premier International Research Center Initiative (WPI Initiative), MEXT, Japan.
\end{acknowledgments}

\appendix{}

\section{one-loop corrections to the quartic coupling coefficients}

Under the parameter assumptions mentioned above, the one-loop corrections to quadratic
and quartic coupling coefficients in the zero temperature potential are given by
\begin{eqnarray}
\Delta m^2_{11}&=&\sum_{i}\frac{ n_i}{64 \pi^2}\alpha^d_{i}\\
\Delta m^2_{22}&=&\sum_{i}\frac{ n_i}{64 \pi^2}\alpha^u_{i}\\
\Delta\lambda_1&=&\sum_{i}\frac{ n_i}{64 \pi^2}\beta^d_{i}+\frac{1}{16 \pi^2}\Bigl \lbrace-3 y_t^4 \frac{\mu^4}{12 M_s^4}\nonumber\\
&&+\sum_{i}N_{C_i}\Bigl [ k^{'4}_i\Bigl (\mathrm{ln}\frac{M^2_S}{M^2_V}-\frac{1}{6}(5-\frac{M^2_V}{M^2_S})(1-\frac{M^2_V}{M^2_S})
+\hat{X}^2_{1}(1-\frac{M^2_V}{3M^2_S})-\frac{\hat{X}^4_{1}}{12}\Bigr)\Bigr ]\Bigr \rbrace\\
\Delta\lambda_2&=&\sum_{i}\frac{ n_i}{64 \pi^2}\beta^u_{i}+\frac{1}{16 \pi^2}\Bigl \lbrace 3 y_t^4 \Bigl[\ln(\frac{\tilde{m}^2_t}{m^2_t})
+\frac{A_t^2}{\tilde{m}^2_t}(1-\frac{A_t^2}{12\tilde{m}^2_t})-\frac{\mu^4}{12\tilde{m}^4_t}\Bigr]\nonumber\\
&&+\sum_{i}N_{C_i}\Bigl [ k^{4}_i\Bigl (\mathrm{ln}\frac{M^2_S}{M^2_V}-\frac{1}{6}(5-\frac{M^2_V}{M^2_S})(1-\frac{M^2_V}{M^2_S})
+\hat{X}^2_{1}(1-\frac{M^2_V}{3M^2_S})-\frac{\hat{X}^4_{1}}{12}\Bigr)\Bigr ]\Bigr \rbrace\\
\Delta\lambda_3&=&\sum_{i}\frac{ n_i}{64 \pi^2}\beta^{ud}_{i}+\frac{1}{16 \pi^2} \Bigl \lbrace3 \Bigl[\frac{y_t^4}{2}\frac{\mu^2}{\tilde{m}^2_t}(1-\frac{A_t^2}{2\tilde{m}^2_t}) \Bigr]\nonumber\\
&&+\sum_{i}N_{C_i}\Bigl [ k^2_ik^{'2}_i\Bigl (-(1-\frac{M^2_V}{M^2_S})^2-\frac{1}{3}(\hat{X}_{1}+\hat{X}_{2})^2 \Bigr )\Bigr ]\Bigr \rbrace\\
\Delta\lambda_6&=&\frac{1}{16 \pi^2}\Bigl \lbrace3 y_t^4\frac{\mu^3A_t}{12\tilde{m}^4_t}\nonumber\\
&&+\sum_{i}N_{C_i}\Bigl [  k^{'3}_ik_i\Bigl (-\frac{2}{3}(2-\frac{M^2_V}{M^2_S})(1-\frac{M^2_V}{M^2_S})-\frac{1}{3}(2\hat{X}^2_{2}+\hat{X}_{1}\hat{X}_{2})\Bigr )\Bigr] \Bigr \rbrace \\
\Delta\lambda_7&=&\frac{1}{16 \pi^2}\Bigl \lbrace3 y_t^4\frac{\mu A_t}{\tilde{m}^2_t} (-\frac{1}{2}+\frac{A_t^2}{12\tilde{m}^2_t})\nonumber\\
&&+\sum_{i}N_{C_i}\Bigl [ k^3_ik^{'}_i\Bigl (-\frac{2}{3}(2-\frac{M^2_V}{M^2_S})(1-\frac{M^2_V}{M^2_S})-\frac{1}{3}(2\hat{X}^2_{1}+\hat{X}_{1}\hat{X}_{2})\Bigr )\Bigr]  \Bigr \rbrace
\end{eqnarray}

\end{document}